\documentclass[11pt,a4paper]{article}
\usepackage{jcappub}
\usepackage{amsbsy,esint,mathrsfs,commath,amsfonts,mathtools,amsthm}
\usepackage[usenames,dvipsnames]{xcolor}
\usepackage{subfig}
\usepackage[english]{babel}
\usepackage{tensor}

\newcommand*\diff{\mathop{}\!\mathrm{d}}

\newcommand{\units}[1]{\; \mathrm{#1}} % unidades sin cursiva y espaciadas

\begin{document}
 
\title{Effects of massive spin-2 fields on gravitational wave propagation}

 \date{\today}

\author[a]{Jose A. R. Cembranos,}
\author[a]{\'Alvaro Cendal}
\author[a]{and Hector Villarrubia-Rojo}
\affiliation[a]{Departamento de F\'isica Te\'orica and IPARCOS, Facultad de Ciencias F\'isicas, Universidad Complutense de Madrid, \\ 
Plaza de las Ciencias 1, 28040 Madrid, Spain}
\emailAdd{cembra@fis.ucm.es}
\emailAdd{acendal@ucm.es}
\emailAdd{hectorvi@ucm.es}

\abstract{
    Massive spin-2 fields in addition to the standard massless graviton arise naturally in extensions of general relativity, such as massive bigravity or models with extra dimensions. This work explores the observational signatures of these fields on the propagation of gravitational waves. Adopting a phenomenological framework consistent with such theories, we derive an analytical transfer function in the ultrarelativistic limit and establish detectability bounds for inspiral sources. Finally, we provide forecasts for the accessible parameter space using current and future gravitational wave detectors.
}

\subheader{\normalfont IPARCOS-UCM-26-002}

\maketitle

\section{Introduction}

The current boom in gravitational wave (GW) astronomy has opened a new window to further test gravity beyond general relativity (GR). After a whole century of experimental successes, the discovery of GWs by the LIGO collaboration in 2015 \cite{LV_GW150914} was yet another successful prediction of GR, and at the same time opened the door to a new family of tests on the theory \cite{Cornish+2011, GairEtAl2013, YunesSiemens2013}. Indeed, modifications of GR often lead to deviations in the properties of GWs, affecting their generation, propagation or polarization \cite{Will1993, EzquiagaZumalacarregui2018}. These tests, applied to the events detected so far by the LIGO-Virgo-KAGRA (LVK) collaboration, have shown no evidence for deviations from GR \cite{LVK_GWTC3_GRTests}. The upcoming generation of GW detectors, such as the Einstein Telescope (ET) \cite{ET2025} and Cosmic Explorer (CE) \cite{CE2021}, as well as the space-based LISA mission \cite{LISA2024}, will further increase the sensitivity of these tests, probing new regimes of gravity.

Following a particle physics approach, GR can be derived as one of the two consistent theories describing a self-interacting massless spin-2 field \cite{Gupta1954, Weinberg1964}, the other being unimodular gravity \cite{BarceloEtAl2014}. A natural extension of these theories is to consider additional massive spin-2 fields. This scenario arises naturally in many extensions to the standard model of particle physics and gravity, such as massive bigravity \cite{HassanRosen2012, LagosFerreira2014,CusinEtAl2015,MaxEtAl2017,BrizuelaEtAl2025}, Randall-Sundrum \cite{RandallSundrum1999} or Kaluza-Klein \cite{MaartensKoyama2010} models. The presence of a mass term explicitly breaks the linear gauge invariance. Additionally, all these theories reduce to the well-known Fierz-Pauli theory in the flat limit, describing a massive, ghost-free spin-2 field in Minkowski spacetime \cite{FierzPauli1939}.

Constructing a consistent non-linear completion of Fierz-Pauli theory is, however, non-trivial. Early attempts to introduce non-linear interactions led to the Boulware-Deser ghost \cite{BoulwareDeser1972}, an additional scalar degree of freedom that renders the theory unstable. These theoretical pathologies were finally overcome by the de Rham-Gabadadze-Tolley (dRGT) massive gravity \cite{RhamEtAl2011} and its bimetric formulation by Hassan and Rosen \cite{HassanRosen2012}. Crucially, these non-linear completions naturally incorporate screening mechanisms, such as the Vainshtein mechanism \cite{Vainshtein1972}. This allows them to overcome the van Dam-Veltman-Zakharov (vDVZ) discontinuity \cite{vanDamVeltman1970,Zakharov1970} ---a consequence of the extra scalar mode that leads to observable deviations from GR even in the massless limit---thereby restoring agreement with general relativity at solar system scales. Full reviews on massive gravity and its extensions can be found in \cite{Hinterbichler2012, Rham2014}.

The main goal of this work is to explore the potential imprints of massive spin-2 fields on gravitational wave propagation, providing forecasts for the detectability of their parameters with current and future detectors. To this end, we adopt a general phenomenological framework, describing the propagation of both massless and massive tensor modes via coupled massive Klein-Gordon equations. While these equations are not derived here from a specific fundamental Lagrangian, they are known to describe the propagation of tensor modes in certain regimes of several well-known massive gravity theories \cite{LagosFerreira2014,CusinEtAl2015,MaxEtAl2017,BrizuelaEtAl2025}. Previous studies in these scenarios have identified an oscillation phenomenon between the massless and massive modes, analogous to neutrino oscillations \cite{MaxEtAl2017, MaxEtAl2018}. In this regard, our model can be understood as a particular case of the more general spin-2 mixing formalism presented in \cite{BeltranJimenezEtAl2020}.

We derive an analytical expression for the waveform transfer function in the ultrarelativistic limit using the WKB approximation. To assess the observability of these effects, we employ the Lindblom distinguishability criterion \cite{LindblomEtAl2008}, implemented through a simplified ``box-like'' detector approximation. This analytical framework allows us to perform computationally efficient detectability forecasts for the current LIGO-Virgo-KAGRA (LVK) network (using the already published catalogs \cite{GWTC4}) as well as for the upcoming third-generation ground-based detectors (Einstein Telescope and Cosmic Explorer) and the space-based LISA mission. To obtain realistic detectability forecasts for these future detectors, we simulate synthetic populations of inspiral gravitational wave events using the \texttt{gwtoolbox} library \cite{GWToolbox}. By doing so, our study extends the work presented in \cite{NarikawaEtAl2015}, which focused on the specific case of bigravity theories and analyzed a limited set of representative events for the LVK network.

The paper is organized as follows. In Section \ref{sec:propagation}, we present the theoretical framework for the propagation of massive tensor modes in a cosmological background, deriving the transfer function in the ultrarelativistic limit via the WKB approximation. Section \ref{sec:detectability_criterion} reviews the Lindblom detectability criterion used to quantify the distinguishability between general relativity and the massive spin-2 waveforms. In Section \ref{sec:analytical_approximation}, we develop the analytical ``box-like'' detector approximation, which allows for efficient computation of detectability bounds. Section \ref{sec:detectability_forecast} presents our main results, providing detectability forecasts for the LVK network (using the GWTC-4 catalog), as well as for future detectors such as ET, CE, and LISA. Finally, Section \ref{sec:conclusions} summarizes our results and conclusions.

\section{Propagation of massless and massive tensor modes}\label{sec:propagation}
Consider tensor perturbations on a flat Friedmann-Lema\^itre-Robertson-Walker (FLRW) spacetime\footnote{Note that there are two main conventions for the definition of tensor perturbations: \textit{i)} the QFT convention, where $h_{ij}$ has dimensions, $[h_{ij}] = L^{-1}$, and \textit{ii)} the GR convention, where $h_{ij}$ is dimensionless. In this work we will use the QFT (dimensionful) convention, which allows us to easily relate linearized GR with Fierz-Pauli theory. The results are not affected by this choice.},
\begin{equation}
    \diff s^2 = a^2(\eta) \left[- \diff \eta^2 + \left(\delta_{ij} + \frac{2}{M_g}  h_{ij} + \mathcal{O}(h^2)\right) \diff x^i \diff x^j \right], \quad \kappa \equiv M_g^{-1} \equiv \sqrt{8 \pi G},
\end{equation} 
where $h_{ij} = \sum_{+,\times} h_A e^A_{ij}$, $A = +, \times$ are the two orthogonal polarization states and $e^A_{ij}$ are the polarization tensors.
At linear order, each Fourier mode $h_k$ of a given polarization obeys the equation of motion
\begin{equation}
    \label{eq:}
    h_k'' + 2 \mathcal{H} h_k' + k^2 h_k = 0,
\end{equation}
where $\mathcal{H} = a'/a$ is the conformal Hubble parameter and primes denote derivatives with respect to conformal time $\eta$.

In this work, we will study the effects of an additional massive tensor mode $\tilde{h}_{ij}$ on the propagation of gravitational waves. The equation of motion for each Fourier mode $\tilde{h}_k$ of mass $m$ is
\begin{equation}
    \tilde{h}_k'' + 2 \mathcal{H} \tilde{h}_k' + (k^2 + a^2 m^2) \tilde{h}_k = 0.
\end{equation}

Introducing the scaled modes $u_k = a h_k$ and $v_k = a \tilde{h}_k$, the equations for the two modes read
\begin{subequations}
    \label{eq:eigenstates_eoms}
    \begin{align}
    u_k'' + \left(k^2 - \frac{a''}{a}\right) u_k &= 0, \\
    v_k'' + \left(k^2 + a^2 m^2 - \frac{a''}{a}\right) v_k &= 0.
    \end{align}
\end{subequations}
In general, both fields may couple to matter. We can parametrize this interaction through a mixing angle $\theta$, such that
\begin{subequations}
    \label{eq:eigenstates_coupled_eoms}
    \begin{align}
        u_k'' + \left(k^2 - \frac{a''}{a}\right) u_k &= \frac{\cos \theta}{M_g} a^3 \Pi_k, \label{eq:eigenstates_coupled_eoms_u} \\
        v_k'' + \left(k^2 + a^2 m^2 - \frac{a''}{a}\right) v_k &= -\frac{\sin \theta}{M_g} a^3 \Pi_k, \label{eq:eigenstates_coupled_eoms_v}
    \end{align}
\end{subequations}
where $\Pi_k$ is the Fourier mode of the anisotropic stress tensor. 

In analogy with neutrino oscillations, we introduce interaction eigenstates $(\mu_k, \tilde{\mu}_k)$ in terms of the mass (or propagation) eigenstates $(u_k, v_k)$ as
\begin{align}
    \begin{pmatrix}
        \mu_k \\
        \tilde{\mu}_k
    \end{pmatrix}
    = 
    \begin{pmatrix}
        \cos \theta & - \sin \theta \\
        \sin \theta & \cos \theta
    \end{pmatrix}
    \begin{pmatrix}
        u_k \\
        v_k
    \end{pmatrix},
\end{align}
where $\mu_k$ is the mode that interacts with matter, while $\tilde{\mu}_k$ is a sterile mode. The equations of motion for the interaction eigenstates read
\begin{subequations}
    \label{eq:interaction_eigenstates_eoms}
    \begin{align}
        \mu_k'' + \left(k^2 - \frac{a''}{a}\right) \mu_k  + a^2 m^2 \sin^2 \theta \big( \mu_k - \cot \theta \, \tilde{\mu}_k \big) &= \frac{1}{M_g} a^3 \Pi_k, \\
        \tilde{\mu}_k'' + \left(k^2 - \frac{a''}{a}\right) \tilde{\mu}_k + a^2 m^2 \cos^2 \theta \big( \tilde{\mu}_k + \tan \theta \, \mu_k \big) &= 0.
    \end{align}
\end{subequations}

Although we will not focus on any fundamental model in this paper, it is worth mentioning that this kind of scenario arises naturally ---typically as an approximate effective description rather than an exact equivalence--- in the context of massive bimetric theories of gravity \cite{HassanRosen2012,LagosFerreira2014, CusinEtAl2015, MaxEtAl2017, BrizuelaEtAl2025}, as well as in Randall-Sundrum \cite{EastherEtAl2003} and Kaluza-Klein \cite{VerasEtAl2015} models. For example, in Hassan-Rosen bigravity, the equivalence between the decoupled modes described in Eq. \eqref{eq:eigenstates_eoms} and the linearized modes of the theory only holds on specific backgrounds (such as Minkowski or de Sitter)\cite{BrizuelaEtAl2025}, but it has been shown to be a good approximation in low-redshift regimes \cite{MaxEtAl2017}.

Moreover, in the flat limit, the linearization of all these theories ---including the one described by \eqref{eq:eigenstates_coupled_eoms}--- reduces to the Fierz-Pauli theory \cite{FierzPauli1939}, which describes a massive, ghost-free spin-2 field $X_{\mu \nu}$ in Minkowski spacetime, with Lagrangian density
\begin{align}
    \label{eq:FP_lagrangian}
    \mathcal{L}_{\text{FP}} (X,m) = &-\frac{1}{2} \partial^{\alpha} X^{\mu \nu}  \big[\partial_{\alpha} X_{\mu \nu} - 2 \partial_{(\mu} X_{\nu) \alpha} - \partial_{\alpha} X \eta_{\mu\nu} + 2 \partial_{(\mu} X \eta_{\nu) \alpha}  \big] \nonumber \\
    &- \frac{1}{2} m^2 (X_{\mu \nu} X^{\mu \nu} - X^2),
\end{align}
where $A_{(\mu \nu)} = \frac{1}{2} (A_{\mu \nu} + A_{\nu \mu})$ and $X = X^{\mu}_{\mu}$.
The full Lagrangian describing the dynamics of both massless and massive spin-2 fields interacting with matter in the flat limit can be expressed as
\begin{align}
    \label{eq:FP_full_lagrangian}
    \mathcal{L} &= \mathcal{L}_{\text{FP}} (h, m = 0) + \mathcal{L}_{\text{FP}} (\tilde{h},m) + \frac{1}{M_g} \left( \cos \theta h_{\mu \nu} + \sin \theta \tilde{h}_{\mu \nu} \right) T^{\mu \nu},
\end{align}
which leads to the flat limit of Eqs.~\eqref{eq:eigenstates_coupled_eoms}.

\subsection{Relating GR and modified coupling constants}
To be able to compare our results with other works in the literature, we will need to relate the coupling $G = 1/(8 \pi M_g^2)$ in our model with the value of the Newtonian gravitational constant \mbox{$G_N = 6.67 \times 10^{-11} \units{m^3 kg^{-1}s^{-2}}$} measured in the laboratory, e.g. in torsion balance experiments. To do so, we will consider the potential generated by a pair of point masses $M_1$ and $M_2$ separated by a distance $r$ in the theory described by the Lagrangian \eqref{eq:FP_full_lagrangian}, which presents a Yukawa-like correction due to the presence of the massive spin-2 field:
\begin{equation}
    \label{eq:yukawa_potential}
    V(r) = -\frac{M_1 M_2 }{r}  \frac{1}{8 \pi M_g^2} \left(\cos^2 \theta + \frac{4}{3} \sin^2 \theta e^{-m r}\right).
\end{equation}
The factor $4/3$ arises from the van Dam-Veltman-Zakharov (vDVZ) discontinuity \cite{vanDamVeltman1970,Zakharov1970}, a consequence of having a scalar mode in the massive theory that leads to observable deviations from GR even in the $m \to 0$ limit. These discontinuities have been widely studied in the context of different massive theories of gravity, and several non-linear mechanisms have been proposed to screen the effects of the scalar mode at small scales, such as the Vainshtein mechanism \cite{Vainshtein1972}. In the following, we assume that our theory presents such a screening mechanism, so that the $4/3$ factor is effectively removed in the small-scale limit relevant for laboratory experiments.\footnote{If we do not consider such a mechanism, $G = G_N (1 + \sin^2(\theta)/3)$. For small couplings, the difference is negligible.} Therefore, the Newtonian gravitational constant $G_N$ is the same as the coupling $G$ appearing in our equations, $G = G_N$.
\subsection{Propagation of interaction eigenstates and transfer function}
We will be interested in studying the interaction eigenstate $h^{\text{int}}_{ij} =\cos \theta \, h_{ij} + \sin \theta \, \tilde{h}_{ij}$ or, equivalently, its scaled version $\mu_{ij} = a h^{\text{int}}_{ij} = \cos \theta u_{ij} + \sin \theta v_{ij}$. This is the mode that interacts with matter, and therefore the one that will be produced in astrophysical processes and detected by GW detectors. We will further assume that the production of gravitational waves in the modified interaction basis $h^{\text{int}}_{ij}$ is the same as in GR, i.e., $h^{\text{int}}_{k}(\eta_s) = h^{\text{GR}}_{k}(\eta_s)$, neglecting the contribution of the additional scalar mode. 

To study the propagation of the interaction eigenstate $h^{\text{int}}_{ij}$ from the source (at $\eta_s$) to the detector (at $\eta_0$), we will apply the WKB (Wentzel-Kramers-Brillouin) approximation, writing the solution to \eqref{eq:eigenstates_eoms} as
\begin{align}
    X_k (\eta_0) & \simeq \frac{C}{\sqrt{2 \omega_X (\eta)}} \, \exp \! \left[-i\int_{\eta_s}^{\eta_0} \omega_X(\eta) \diff \eta\right],
\end{align}
where $X = u, v$ and $\omega_u^2(\eta) = k^2$, $\omega_v^2(\eta) = k^2 + a^2 (\eta) m^2$. Fixing the initial conditions $h^{\text{int}}_{k}(\eta_s) = h^{\text{GR}}_{k}(\eta_s)$, we find
\begin{align}
    u_k (\eta_0) & \simeq \cos \theta \mu^{\text{GR}}_k(\eta_0), \\
    v_k (\eta_0) & \simeq \sin \theta \mu^{\text{GR}}_k(\eta_0) \left(\frac{\omega_v(\eta_s)}{\omega_v(\eta_0)}\right)^{1/2} \, \exp \! \left[-i\int_{\eta_s}^{\eta_0} \big(\omega_v(\eta) - k\big) \diff \eta\right],
\end{align}
where we have explicitly factored out the propagation of the GR mode, 
\begin{equation}
    \mu^{\text{GR}}_k(\eta_0) = \mu^{\text{GR}}_k(\eta_s) \exp[-i k (\eta_0 - \eta_s)],
\end{equation}
which coincides with the propagation of the massless eigenstate $u_k$. This factorization isolates the effects of the modified propagation from the standard GR evolution, and allows us to define a transfer function $F(k, z_s)$ for a source at redshift $z_s = 1/a_s - 1$, relating the modified interaction eigenstate at the detector to the unmodified GR waveform:
\begin{align}
    F(k, z_s) =  \frac{h_k^{\text{int}}(\eta_0)}{h_k^{\text{GR}}(\eta_0)} = \frac{\mu_k}{\mu_k^{\text{GR}}} = \frac{1}{1 + \tan^2 \theta} \left(1 + \mathcal{A}(k, z_s) \tan^2 \theta  \, \exp \! \left[-i \varphi(k, z_s)\right] \right),
\end{align}
where we define the amplitude $\mathcal{A}(k, z_s)$ and phase $\varphi(k, z_s)$ as
\begin{align}
    \mathcal{A}(k, z_s) &= \left(\frac{\omega_v(\eta_s)}{\omega_v(\eta_0)}\right)^{1/2}  = \left(\frac{k^2 + (1+z_s)^{-2} m^2}{k^2 + m^2}\right)^{1/4}, \\
    \varphi(k, z_s) &= \int_{\eta_s}^{\eta_0} \big(\omega_v(\eta) - k\big) \diff \eta  = \int^{z_s}_0 \frac{\sqrt{k^2 + m^2 (1 + z)^{-2}} - k}{H(z)} \diff z.
\end{align}

In the non-relativistic limit ($m \gg k$), the modified propagation results in a strongly suppressed echo, widely separated from the main signal. We therefore focus on the ultrarelativistic limit ($k \gg m$), where the amplitude and phase can be approximated as
\begin{align}
    \mathcal{A}_k &= 1 + \mathcal{O}\left(\frac{m^2}{k^2}\right), \\
    \varphi_k(z_s) &= \frac{m^2}{2k}\int^{z_s}_0 \frac{1}{(1 + z)^2 H(z)}  \diff z + \mathcal{O}\left(\frac{m^4}{k^3}\right).
\end{align}
This approximation is detector-dependent, since the ultrarelativistic condition $k \gg m$ must hold across the full frequency range of the detector. For Earth-based detectors (LVK, ET, CE), the ultrarelativistic limit is valid for $m \ll 10^{-14} \units{eV}$, while for LISA it requires $m \ll 10^{-19} \units{eV}$. To ensure the validity of the approximation for all detectors, we will therefore consider $m < 10^{-20} \units{eV}$.

\begin{figure}
    \centering
    \includegraphics[width=0.75\textwidth]{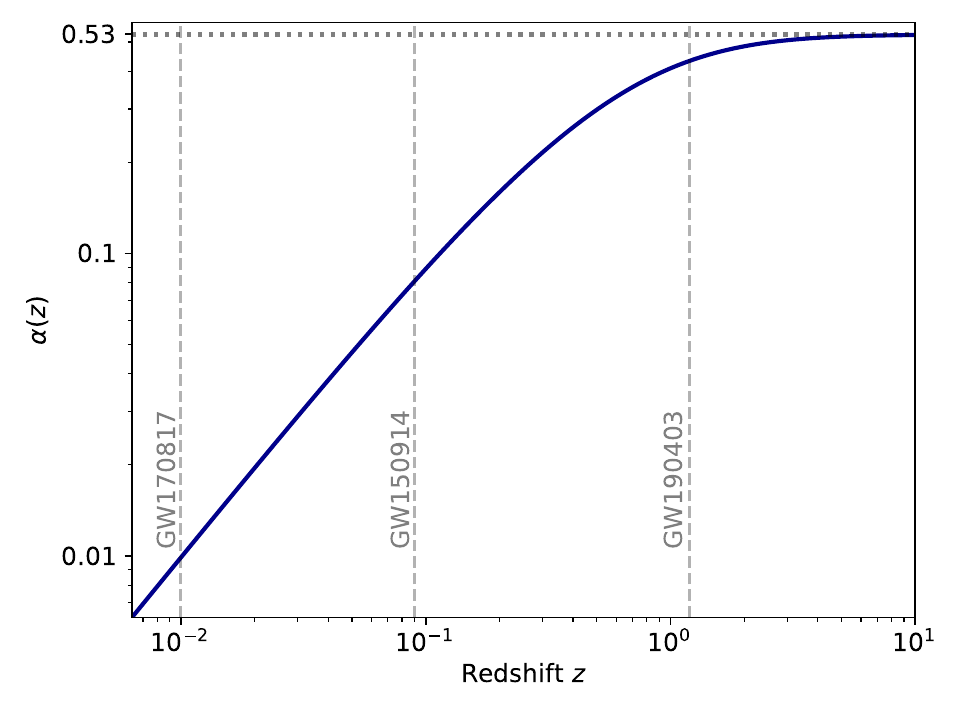}
    \caption{Function $\alpha(z)$ as a function of the redshift $z$. Vertical lines indicate the redshifts of GW170817 (the closest event recorded), GW150914, and GW190403 (the farthest event recorded) for reference.}
    \label{fig:alpha}
\end{figure}

In the following, we focus on GW signals from compact binary coalescences (CBC) and assume a flat $\Lambda \text{CDM}$ universe with cosmological parameters taken from the Planck 18 results \cite{Planck2018}: $\Omega_M = 0.311$, $\Omega_\Lambda = 0.689$, and $H_0 = 67.66 \units{km/s/Mpc}$. In this case, the Hubble parameter can be approximated as $H(z) \simeq H_0 \sqrt{\Omega_M (1 + z)^3 + \Omega_\Lambda}$, where $\Omega_M$ and $\Omega_\Lambda$ are the matter and dark energy density parameters, respectively. The corresponding phase can then be expressed as
\begin{align}
    \label{eq:phase_ultrarel}
    \varphi_k(k, z_s) &= \alpha(z_s) \frac{m^2}{2k H_0}
\end{align}
with the function $\alpha(z_s)$ defined by
\begin{align}
    \alpha(z_s) &= \frac{1}{\sqrt{\Omega_\Lambda}}  \left( \tensor[_{2}]{F}{_{1}}\left[-\frac1 3, \frac1 2;  \frac2 3; - \frac{\Omega_M}{\Omega_{\Lambda}} \right] \quad - \tensor[_{2}]{F}{_{1}}\left[-\frac1 3, \frac1 2;  \frac2 3; - \frac{(1+z_s)^3 \Omega_M}{\Omega_{\Lambda}} \right] (1+z_s)^{-1} \right),
\end{align}
where $\tensor[_{2}]{F}{_{1}}$ is the Gaussian hypergeometric function. In Figure \ref{fig:alpha} we show $\alpha(z)$. For large redshifts, $\alpha(z_s) \to 0.53$, while for $z = 0.01$ ---the redshift of GW170817, the closest event ever detected--- $\alpha(z) \simeq 0.01$, roughly one order of magnitude smaller.

Equation \eqref{eq:phase_ultrarel} introduces a new frequency-dependent mass scale $m_{\varphi}(k) = \sqrt{H_0 k / \alpha}$ where phase modifications become significant. Since the gravitational wave detectors considered in this work operate in the subhorizon regime ($k \gg H_0$), this scale satisfies the hierarchy $H_0 \ll m_{\varphi} \ll k$. For each detector, we can define two characteristic mass scales corresponding to the maximum and minimum frequencies probed by the detector,
\begin{equation}
    \label{eq:phi_scales}
    m_{\varphi}^{\text{max}} = \sqrt{\frac{H_0 2 \pi f_{\text{max}}}{\alpha}}, \quad \quad m_{\varphi}^{\text{min}} = \sqrt{\frac{H_0 2 \pi f_{\text{min}}}{\alpha}},
\end{equation} 
where $f_{\text{max}}$ and $f_{\text{min}}$ are the maximum and minimum frequencies probed by the experiment. 
These scales will be relevant in the following section, determining the different regimes of detectability.

\section{Detectability criterion} \label{sec:detectability_criterion}
To obtain a first estimate of the detectability of the massive spin-2 field effects on gravitational wave observations, we will use the Lindblom criterion \cite{LindblomEtAl2008}. This criterion is based on the mismatch between two waveforms $h_1$ and $h_2$, defined as
\begin{align}
\label{eq:mismatch}
    \mathcal{M} \equiv 1 - \frac{(h_1 | h_2)}{\sqrt{(h_1 | h_1)(h_2 | h_2)}},
\end{align}
where $(h_1 | h_2)$ is the noise-weighted inner product, defined as
\begin{align}
\label{eq:inner_product}
    (h_1 | h_2) = 4 \Re \int^{\infty}_{0} \frac{\tilde{h}_1^*(f) \tilde{h}_2(f)}{S_n(f)} \diff f,
\end{align}
where $S_n(f)$ is the one-sided power spectral density (PSD) of the detector strain noise. From this definition, the signal-to-noise ratio (SNR) of any waveform $h$ can be computed as $\mathrm{SNR}^2 = (h | h)$. In our case, we will consider $h_{1} = h^{\text{GR}}$ ---the original general relativity waveform--- and $h_{2} = h^{\text{int}}$, the modified waveform, that is, $h^{\text{int}}_k = F_k h^{\text{GR}}_k$. 

The Lindblom criterion states that the two waveforms $h_1$ and $h_2$ are indistinguishable if $(\delta h | \delta h) \leq 1$, where $\delta h = h_2 - h_1$. For waveforms with similar SNR, the Lindblom criterion requires $\mathcal{M} < \mathrm{SNR}^{-2}$. Note that the converse of the Lindblom criterion is not necessarily true, i.e., if $\mathcal{M} > \mathrm{SNR}^{-2}$, the two waveforms are not necessarily distinguishable. There could be non-trivial degeneracies between the parameters of the two waveforms, resulting in a lower distinguishability than expected from the (inverse) Lindblom criterion. A full analysis of the parameter space and the degeneracies between the parameters is beyond the scope of this paper. We will consider as the bound of detectability the case where the Lindblom criterion is saturated, i.e., $\mathcal{M} = \mathrm{SNR}^{-2}$. This saturated criterion, together with the assumption of no degeneracies between parameters and full knowledge of the non-modified waveform, will give us the most optimistic bounds on the detectable region of the parameter space.

\section{Analytical approach: box-like detector approximation} \label{sec:analytical_approximation}

Given the criterion $\mathcal{M} = \mathrm{SNR}^{-2}$, the curve of detectability in the $(m, G_h)$ parameter space can be computed numerically for each event and detector. However, to perform a more efficient computation of these bounds over large populations of events, it is useful to derive an analytical approximation to Eq. \eqref{eq:mismatch}.

To do so, we will consider a box-like detector ---that is, a detector with a constant PSD within a frequency range $[f_{\text{min}}, f_{\text{max}}]$ and null sensitivity outside this range: 
\begin{equation}
    \label{eq:box_detector}
    S_n(f) = \begin{cases}
        S, & f \in [f_{\text{min}}, f_{\text{max}}],\\
        \infty, & \text{otherwise}.
    \end{cases}
\end{equation}
We will also restrict our analysis to inspiral sources, assuming that the frequency band of the approximation \eqref{eq:box_detector} covers the inspiral phase of waveform $h^{\text{GR}}$, so it can be approximated by $h^{\text{GR}}(f) \simeq h_0 \, f^{-7/6}$, where $h_0$ is a constant.

Considering the ultrarelativistic limit of the phase \eqref{eq:phase_ultrarel}, we could further simplify the analysis by focusing on the regimes $m \gg m_{\varphi}^{\text{max}}$ or $m \ll m_{\varphi}^{\text{min}}$, where the scales $m_{\varphi}^{\text{max, min}}$ are defined in Eq. \eqref{eq:phi_scales}. In practice, however, we observe that for a given detector the sensitivity is effectively governed by a single mass scale, namely $m_c$. 

This can be understood by inspecting the inner product \eqref{eq:inner_product}. An inspiral waveform (that is, $h(f) \propto f^{-7/6}$) introduces a factor $f^{-7/3}$ in the integrand. This strongly weights the inner product toward low frequencies and suppresses contributions from high frequencies. We will therefore define the characteristic frequency of the detector as the one that maximizes the integrand of the inner product \eqref{eq:inner_product} for an inspiral waveform:
\begin{equation}
    f_c :=  \operatorname*{arg\,max}_f \left( \frac{f^{-7/3}}{S_n(f)} \right).
\end{equation} 
As we will observe later, this characteristic frequency $f_c$ is enough to describe the detectability bounds of the detector. We can define the corresponding characteristic mass scale as
\begin{equation}
    m_c = \sqrt{H_0 2 \pi f_c / \alpha},
\end{equation} 
which will allow us to separate the two regimes of detectability: $m \gg m_c$ and $m \ll m_c$.

Using the approximations introduced earlier ---the box-like detector \eqref{eq:box_detector} and the inspiral waveform $h^{\text{GR}}(f) \simeq h_0 f^{-7/6}$--- we can obtain analytical expressions for the mismatch \eqref{eq:mismatch} in the limits $m \gg m_c$ and $m \ll m_c$. Applying the Lindblom criterion, $\mathcal{M} = \mathrm{SNR}^{-2}$, we obtain the detectability bound
\begin{equation}
    \label{eq:analytical_approximation}
    \tan \theta_{\text{det}}(m) = 
    \begin{cases}
        \left[\frac{2 \text{SNR}^2 - 1}{(\text{SNR}^2 - 1)^2}\right]^{1/4}, &  m \gg m_{c},\\[10pt]
        \frac{1}{ \left[(m/m_{\text{low}})^2 \; - \; 1\right]^{1/2}}, & m_{\text{low}} < m \ll m_{c},
    \end{cases}
\end{equation}
where $m_{\text{low}} = m_c / (\sqrt{5} \, \mathrm{SNR})$. In order to plot a smooth analytical approximation, we take $\tan \theta_{\det} (m)$ as the square root of the sum of the squares of the two limits. For large masses ($m \gg m_c$), the approximation is independent of the spin-2 field mass $m$, depending only on the signal-to-noise ratio. For small masses, the mass scale $m_{\text{low}}$ sets a lower threshold below which the effects of the massive spin-2 field cannot be detected, regardless of the coupling $\tan \theta$.

\begin{figure*}[ht]
    \centering
    \subfloat[LVK, $z = 0.01$, varying total mass]{
        \includegraphics[width=0.50\textwidth]{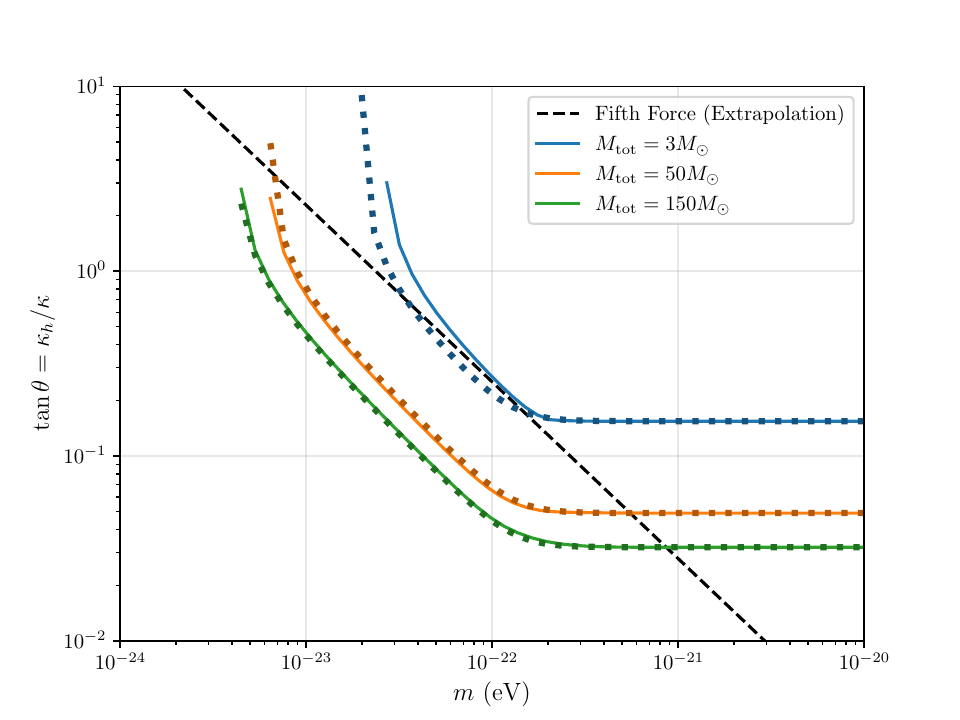}
    }
    \subfloat[LISA, $M_{\text{tot}} = 10^6 M_{\odot}$, varying redshift]{
        \includegraphics[width=0.50\textwidth]{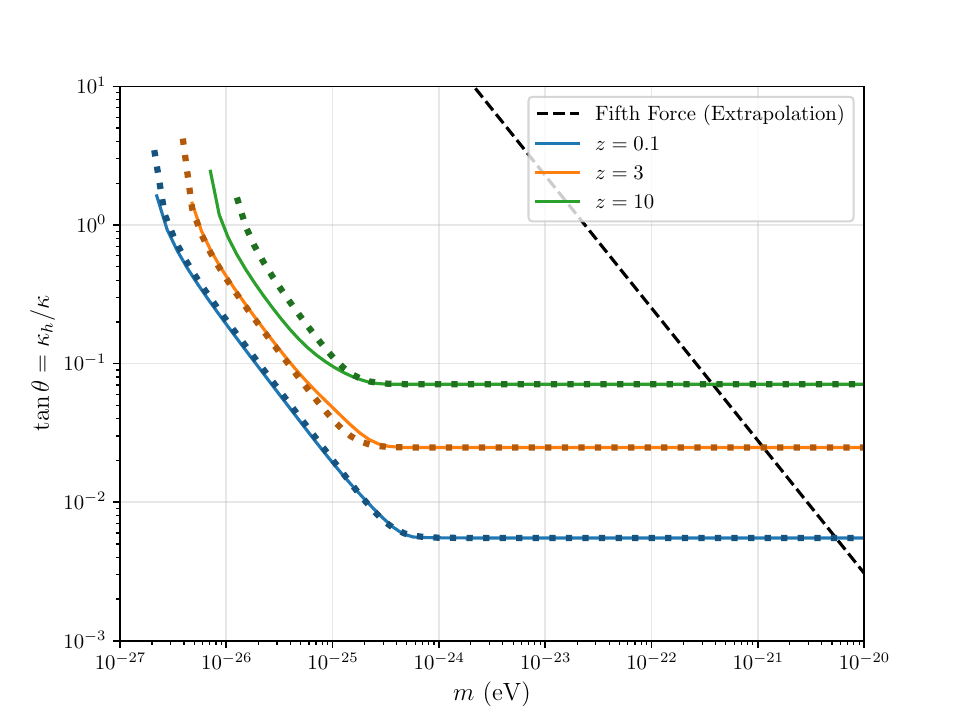}        
    }\\
    \caption{Detectability bounds for different test events, computed both numerically (solid) and using the analytical approximation \eqref{eq:analytical_approximation} (dotted).}
    \label{fig:approx_numerical}
\end{figure*}

Therefore, for a specific detector, given the SNR of the original signal and the redshift of the source, we can compute the approximation $\tan \theta_{\text{det}}(m)$. Figure \ref{fig:approx_numerical} compares the analytically derived detectability bounds \eqref{eq:analytical_approximation} with those obtained from numerical calculations for both LVK and LISA, across a range of source redshifts and total masses. Despite the simplicity of the box-like approximation \eqref{eq:analytical_approximation}, it captures the essential phenomenology of the signal detection. As shown in Fig. \ref{fig:approx_numerical}, the analytical curves (dotted) track the numerical results (solid) with remarkable accuracy across orders of magnitude in mass. Slight deviations appear only at low masses. For the scope of this work, this approximation provides a sufficiently accurate and robust estimate of the detectability.

\section{Detectability forecast} \label{sec:detectability_forecast}
Having established the reliability of the box-like approximation \eqref{eq:analytical_approximation}, we now use it to forecast the detectability of massive spin-2 modes ---described by \eqref{eq:eigenstates_eoms}--- in gravitational-wave observations from current and future detectors. The key advantage of this analytical approximation is that, for a given detector, the detectability bound depends only on the event's SNR and redshift (or equivalently, on $\alpha(z)$). This eliminates the need for a full numerical analysis of each event, a task that would be computationally expensive when considering extensive catalogs of events or simulations.

In this work, we consider three detectors: the current LIGO-Virgo-KAGRA (LVK) network, the planned Einstein Telescope (ET) and Cosmic Explorer (CE) network (ET+CE), and the upcoming space-based LISA mission. For LVK, we give a forecast based on the 174 events with measured redshift reported in GWTC-4 \cite{GWTC4}. For ET+CE and LISA, and to explore the potential future reach of LVK, we generate simulated event populations using the Python package \texttt{gwtoolbox} \cite{GWToolbox}. This tool provides estimates of detectable events by each detector over their respective observing runs (assumed to be 2 years for LIGO and ET+CE and 4.5 years for LISA), as well as simulations of their properties, including the redshifts and SNRs needed to apply the analytical approximation \eqref{eq:analytical_approximation}. For each event in a run, we compute the detectability bound using \eqref{eq:analytical_approximation}. We take the lowest bound for each mass in a simulation, obtaining in this way the overall detectability bound for that run. Given that each simulation corresponds to a different realization of the population of events, we run hundreds of simulations for each detector, obtaining a distribution of detectability bounds for each detector.

The results are shown in Fig.~\ref{fig:detectability_forecast}. In order to avoid model-dependence, we have chosen to compare the detectability bounds obtained in this work with the constraints from \cite{CembranosEtAl2017,CembranosEtAl2022}, which only assume the presence of a massive spin-2 field coupled to matter as in \eqref{eq:FP_full_lagrangian}, without further assumptions on the non-linear completion of the theory. From Fig.~\ref{fig:detectability_forecast} we see that the range of masses where the detectability bounds from GW observations improve over previous constraints is $m \ll 10^{-20} \units{eV}$, well below the ultrarelativistic limit considered in this work. 

\begin{figure}
    \centering
    \includegraphics[width=0.9\textwidth]{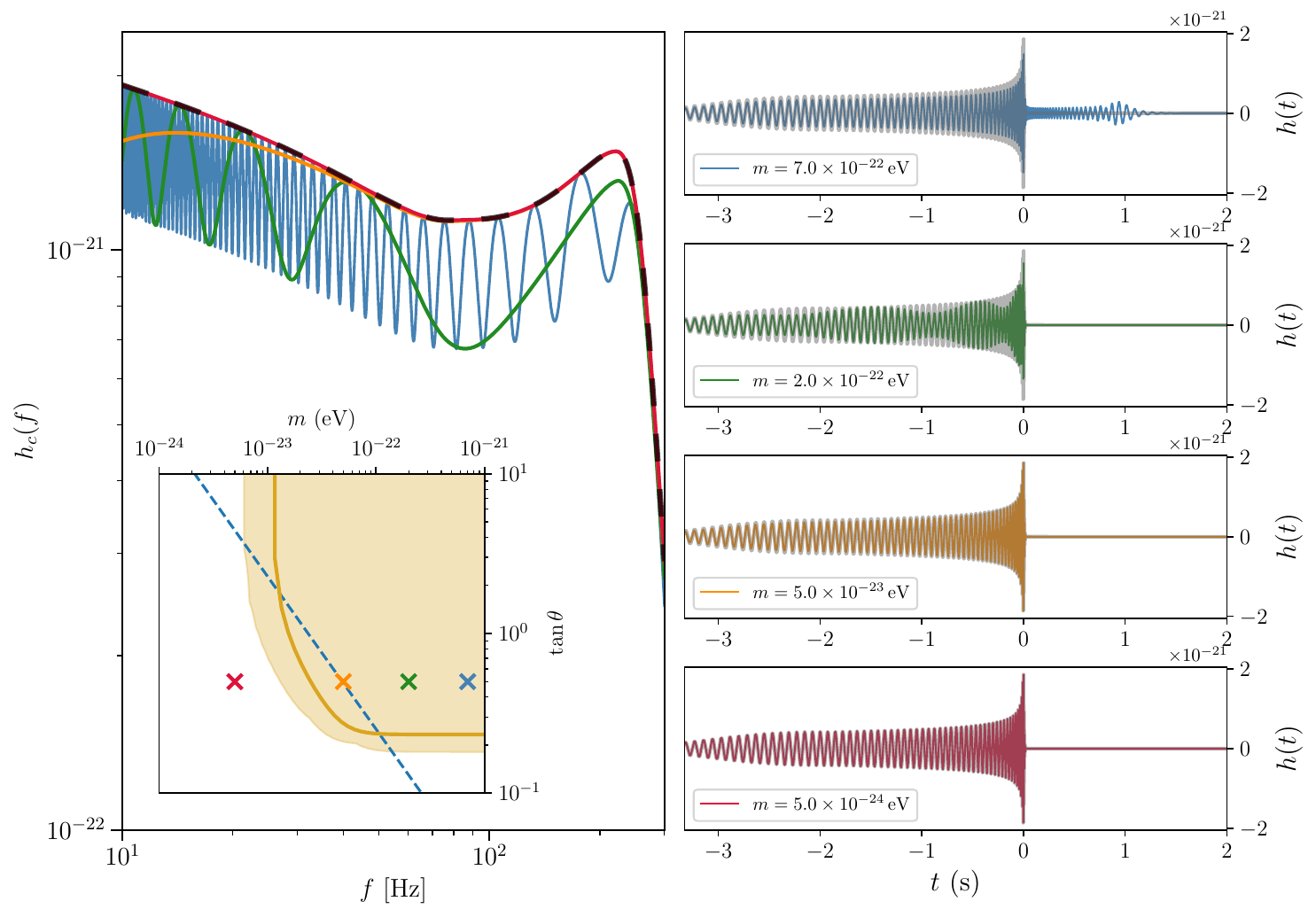}
    \caption{Examples of the effects of the massive spin-2 field on GW150914-like waveforms for different values of the mass $m$, fixing $\tan \theta = 0.5$, both in the frequency (left) and time (right) domains. The black line corresponds to the original GR waveform. Inset shows the position of $(m, \tan \theta)$ in the detectability forecast of LVK from GWTC-4 events.}
    \label{fig:waveforms}
\end{figure}

To further illustrate the effects of the massive spin-2 field on gravitational waveforms, we present in Figure \ref{fig:waveforms} examples of modified waveforms (based on GW150914) for different values of the mass $m$, fixing the coupling parameter $\tan \theta = 0.5$, both in the frequency and time domains.

\section{Summary and conclusions} \label{sec:conclusions}

The results presented in this work demonstrate the potential and limitations of current and future gravitational wave detectors to probe the effects of massive tensor modes on the propagation of gravitational waves. Using a phenomenological framework consistent with the linearized dynamics of theories such as massive bigravity, we have derived an analytical approximation for the detectability bounds in the parameter space of mass $m$ and coupling $\tan \theta$ (Eq. \eqref{eq:analytical_approximation}) for the case of inspiral sources. This approximation proves to be accurate when compared to numerical calculations, allowing us to efficiently forecast the detectability of these effects for different detectors and event populations.

\begin{figure}
    \centering
    \subfloat[Low mass region. The hatched area indicates the 95\% confidence region for the lower mass detectability bound of each detector, obtained from simulated event populations. For LVK, the results from GWTC-4 events are also shown (solid gold region).]{
        \includegraphics[width=0.75\textwidth]{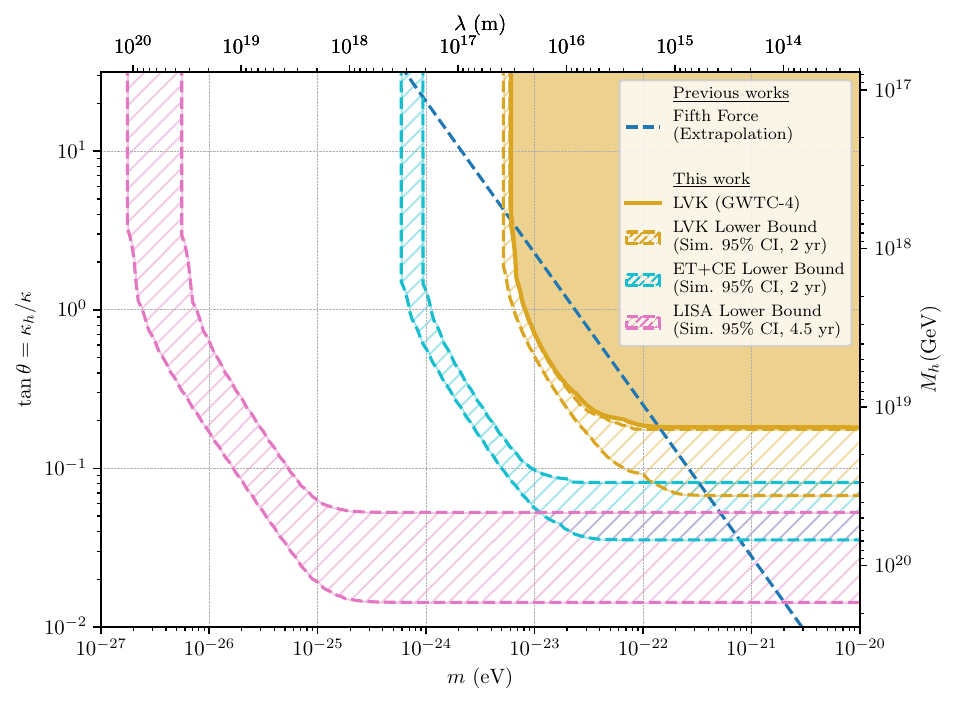}
    }\\
    \subfloat[Comparison of the detectability bounds obtained in this work with previous constraints. For LVK, the results from GWTC-4 events are shown (solid gold region). For ET+CE and LISA, the median from the simulated event populations is shown (solid light blue and pink regions, respectively).]{
        \includegraphics[width=0.75\textwidth]{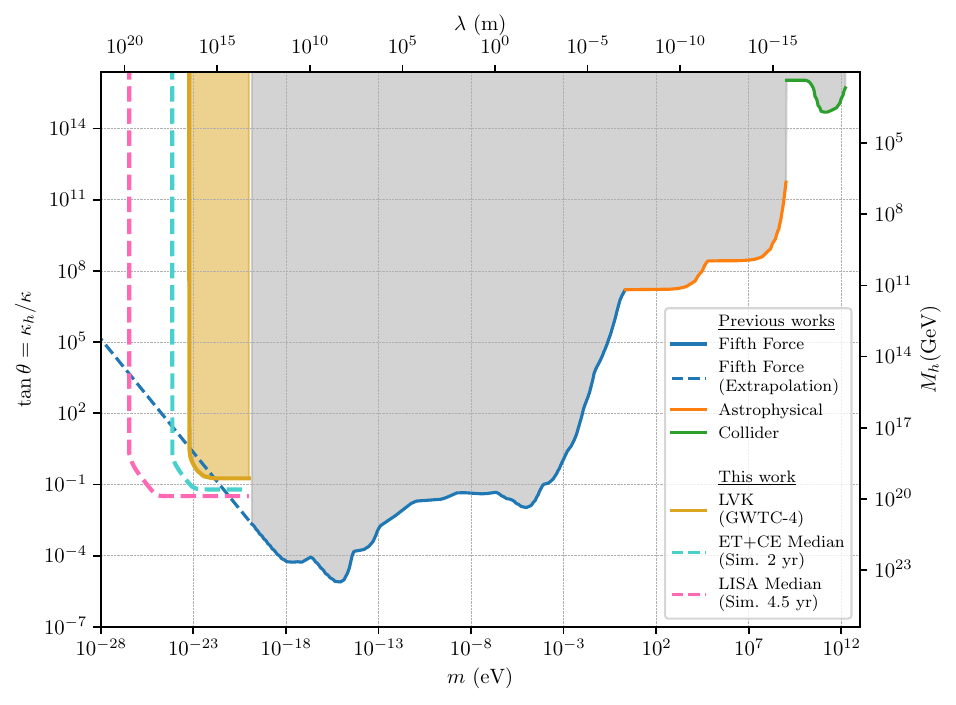}
    }
    \caption{Detectability forecast for the LVK network (gold), ET+CE network (light blue), and LISA (pink).}
    \label{fig:detectability_forecast}
\end{figure}

The results shown in Fig.~\ref{fig:detectability_forecast} represent the detectability bounds obtained from the Lindblom criterion (see Sec.~\ref{sec:detectability_criterion}). These bounds are not meant to be interpreted as excluded or constrained regions of the parameter space, something that would require a full case-by-case study taking into consideration degeneracies and other subtleties.\footnote{For an example of such an analysis, see the study of GW150914 and GW151226 in \cite{MaxEtAl2017}.} Instead, they indicate the regions where the effects of the massive spin-2 field on gravitational wave signals would be detectable assuming a full knowledge of the original waveform, which is the most optimistic scenario. 
In this situation, any effects on the propagation of GWs due to massive modes with $(m, \tan \theta)$ under the curves of Fig.~\ref{fig:detectability_forecast} are undetectable according to the Lindblom criterion, even considering the most optimistic scenario. Thus, those curves show the maximum reach of each detector in probing the effects of massive spin-2 fields on gravitational wave propagation. 

In summary, we have presented the detectability forecasts for the effects of massive spin-2 fields on gravitational wave propagation for the main current and future GW detectors, establishing a lower bound on the accessible parameter space. These results identify the most promising regions of the parameter space to prioritize in future investigations.

\begin{acknowledgments}
This work is partially supported by the projects PID2022-139841NB-I00 and PID2022-138263NB-I00 funded by MICIU/AEI/10.13039/501100011033 and by ERDF/EU. This work is also part of the COST (European Cooperation in Science and Technology) Actions CA21106, CA21136, CA22113 and CA23130. Additionally, Á.C. acknowledges financial support from MIU (Ministerio de Universidades, Spain) fellowship FPU22/03222.
\end{acknowledgments}

\bibliographystyle{JCAP} 
\bibliography{bibliography}

\providecommand{\href}[2]{#2}\begingroup\raggedright\begin{thebibliography}{10}

\bibitem{LV_GW150914}
{\scshape LIGO Scientific Collaboration and Virgo Collaboration} collaboration,
  \emph{Observation of gravitational waves from a binary black hole merger},
  \href{https://doi.org/10.1103/PhysRevLett.116.061102}{\emph{Physical Review
  Letters} {\bfseries 116} (2016) 61102}.

\bibitem{Cornish+2011}
N.~Cornish, L.~Sampson, N.~Yunes and F.~Pretorius, \emph{Gravitational wave
  tests of general relativity with the parameterized post-einsteinian
  framework}, \href{https://doi.org/10.1103/PhysRevD.84.062003}{\emph{Phys.
  Rev. D} {\bfseries 84} (2011) 062003}.

\bibitem{GairEtAl2013}
J.R.~Gair, M.~Vallisneri, S.L.~Larson and J.G.~Baker, \emph{Testing general
  relativity with low-frequency, space-based gravitational-wave detectors},
  \href{https://doi.org/10.12942/lrr-2013-7}{\emph{Living Rev. Rel.} {\bfseries
  16} (2013) 7}.

\bibitem{YunesSiemens2013}
N.~Yunes and X.~Siemens, \emph{Gravitational wave tests of general relativity
  with ground-based detectors and pulsar timing-arrays},
  \href{https://doi.org/10.12942/lrr-2013-9}{\emph{Living Rev. Rel.} {\bfseries
  16} (2013) 9}.

\bibitem{Will1993}
C.M.~Will, \emph{Theory and Experiment in Gravitational Physics}, Cambridge
  University Press, Cambridge, revised~ed. (1993).

\bibitem{EzquiagaZumalacarregui2018}
J.M.~Ezquiaga and M.~Zumalac{\'a}rregui, \emph{Dark energy in light of
  multi-messenger gravitational-wave astronomy},
  \href{https://doi.org/10.3389/fspas.2018.00044}{\emph{Frontiers in Astronomy
  and Space Sciences} {\bfseries 5} (2018) 44}
  [\href{https://arxiv.org/abs/1807.09241}{{\ttfamily 1807.09241}}].

\bibitem{LVK_GWTC3_GRTests}
R.~Abbott, H.~Abe, F.~Acernese, K.~Ackley, N.~Adhikari, R.X.~Adhikari et~al.,
  \emph{Tests of general relativity with {{GWTC-3}}},
  \href{https://doi.org/10.1103/PhysRevD.112.084080}{\emph{Physical Review D}
  {\bfseries 112} (2025) 084080}.

\bibitem{ET2025}
A.~Abac and others {(Einstein Telescope Collaboration)}, \emph{The science of
  the {Einstein Telescope}},  Tech. Rep. arXiv (Mar., 2025).

\bibitem{CE2021}
M.~Evans and others {(Cosmic Explorer Collaboration)}, \emph{A horizon study
  for {Cosmic Explorer}: Science, observatories, and community},  Technical
  Report CE-P2100003-v7 (Oct., 2021).

\bibitem{LISA2024}
M.~Colpi and others {(LISA Consortium)}, \emph{{LISA} definition study report},
   Tech. Rep. ESA-SCI-DIR-RP-002, arXiv (Feb., 2024).

\bibitem{Gupta1954}
S.N.~Gupta, \emph{Gravitation and electromagnetism},
  \href{https://doi.org/10.1103/PhysRev.96.1683}{\emph{Physical Review}
  {\bfseries 96} (1954) 1683}.

\bibitem{Weinberg1964}
S.~Weinberg, \emph{Photons and gravitons in \${{S}}\$-matrix theory:
  {{Derivation}} of charge conservation and equality of gravitational and
  inertial mass},
  \href{https://doi.org/10.1103/PhysRev.135.B1049}{\emph{Physical Review}
  {\bfseries 135} (1964) B1049}.

\bibitem{BarceloEtAl2014}
C.~Barcel{\'o}, R.~{Carballo-Rubio} and L.J.~Garay, \emph{Unimodular gravity
  and general relativity from graviton self-interactions},
  \href{https://doi.org/10.1103/PhysRevD.89.124019}{\emph{Physical Review D}
  {\bfseries 89} (2014) 124019}
  [\href{https://arxiv.org/abs/1401.2941}{{\ttfamily 1401.2941}}].

\bibitem{LagosFerreira2014}
M.~Lagos and P.G.~Ferreira, \emph{Cosmological perturbations in massive
  bigravity}, \href{https://doi.org/10.1088/1475-7516/2014/12/026}{\emph{JCAP}
  {\bfseries 12} (2014) 026}.

\bibitem{CusinEtAl2015}
G.~Cusin, R.~Durrer, P.~Guarato and M.~Motta, \emph{Gravitational waves in
  bigravity cosmology},
  \href{https://doi.org/10.1088/1475-7516/2015/05/030}{\emph{Journal of
  Cosmology and Astroparticle Physics} {\bfseries 2015} (2015) 30}
  [\href{https://arxiv.org/abs/1412.5979}{{\ttfamily 1412.5979}}].

\bibitem{MaxEtAl2017}
K.~Max, M.~Platscher and J.~Smirnov, \emph{Gravitational wave oscillations in
  bigravity},
  \href{https://doi.org/10.1103/PhysRevLett.119.111101}{\emph{Physical Review
  Letters} {\bfseries 119} (2017) 111101}
  [\href{https://arxiv.org/abs/1703.07785}{{\ttfamily 1703.07785}}].

\bibitem{BrizuelaEtAl2025}
D.~Brizuela, M.~de~Cesare and A.S.~Oficial, \emph{Gravitational wave
  propagation in bigravity in the late universe},  July, 2025.
\newblock 10.48550/arXiv.2507.11526.

\bibitem{RandallSundrum1999}
L.~Randall and R.~Sundrum, \emph{A large mass hierarchy from a small extra
  dimension}, \href{https://doi.org/10.1103/PhysRevLett.83.3370}{\emph{Physical
  Review Letters} {\bfseries 83} (1999) 3370}
  [\href{https://arxiv.org/abs/hep-ph/9905221}{{\ttfamily hep-ph/9905221}}].

\bibitem{MaartensKoyama2010}
R.~Maartens and K.~Koyama, \emph{Brane-world gravity},
  \href{https://doi.org/10.12942/lrr-2010-5}{\emph{Living Reviews in
  Relativity} {\bfseries 13} (2010) 5}.

\bibitem{FierzPauli1939}
M.~Fierz and W.~Pauli, \emph{On relativistic wave equations for particles of
  arbitrary spin in an electromagnetic field},
  \href{https://doi.org/10.1098/rspa.1939.0140}{\emph{Proc. R. Soc. London,
  Ser. A} {\bfseries 173} (1939) 211}.

\bibitem{BoulwareDeser1972}
D.G.~Boulware and S.~Deser, \emph{Can gravitation have a finite range?},
  \href{https://doi.org/10.1103/PhysRevD.6.3368}{\emph{Physical Review D}
  {\bfseries 6} (1972) 3368}.

\bibitem{RhamEtAl2011}
C.~de~Rham, G.~Gabadadze and A.J.~Tolley, \emph{Resummation of massive
  gravity},
  \href{https://doi.org/10.1103/PhysRevLett.106.231101}{\emph{Physical Review
  Letters} {\bfseries 106} (2011) 231101}
  [\href{https://arxiv.org/abs/1011.1232}{{\ttfamily 1011.1232}}].

\bibitem{HassanRosen2012}
S.F.~Hassan and R.A.~Rosen, \emph{Bimetric gravity from ghost-free massive
  gravity}, \href{https://doi.org/10.1007/JHEP02(2012)126}{\emph{Journal of
  High Energy Physics} {\bfseries 2012} (2012) 126}
  [\href{https://arxiv.org/abs/1109.3515}{{\ttfamily 1109.3515}}].

\bibitem{Vainshtein1972}
A.~Vainshtein, \emph{To the problem of nonvanishing gravitation mass},
  \href{https://doi.org/10.1016/0370-2693(72)90147-5}{\emph{Physics Letters B}
  {\bfseries 39} (1972) 393}.

\bibitem{vanDamVeltman1970}
H.~{van Dam} and M.~Veltman, \emph{Massive and mass-less {{Yang-Mills}} and
  gravitational fields},
  \href{https://doi.org/10.1016/0550-3213(70)90416-5}{\emph{Nuclear Physics B}
  {\bfseries 22} (1970) 397}.

\bibitem{Zakharov1970}
V.I.~Zakharov, \emph{Linearized gravitation theory and the graviton mass},
  {\emph{JETP Lett.} {\bfseries 12} (1970) 312}.

\bibitem{Hinterbichler2012}
K.~Hinterbichler, \emph{Theoretical aspects of massive gravity},
  \href{https://doi.org/10.1103/RevModPhys.84.671}{\emph{Reviews of Modern
  Physics} {\bfseries 84} (2012) 671}
  [\href{https://arxiv.org/abs/1105.3735}{{\ttfamily 1105.3735}}].

\bibitem{Rham2014}
C.~de~Rham, \emph{Massive gravity},
  \href{https://doi.org/10.12942/lrr-2014-7}{\emph{Living Reviews in
  Relativity} {\bfseries 17} (2014) 7}
  [\href{https://arxiv.org/abs/1401.4173}{{\ttfamily 1401.4173}}].

\bibitem{MaxEtAl2018}
K.~Max, M.~Platscher and J.~Smirnov, \emph{Decoherence of gravitational wave
  oscillations in bigravity},
  \href{https://doi.org/10.1103/PhysRevD.97.064009}{\emph{Physical Review D}
  {\bfseries 97} (2018) 064009}
  [\href{https://arxiv.org/abs/1712.06601}{{\ttfamily 1712.06601}}].

\bibitem{BeltranJimenezEtAl2020}
J.~Beltr{\'a}n~Jim{\'e}nez, J.M.~Ezquiaga and L.~Heisenberg, \emph{Probing
  cosmological fields with gravitational wave oscillations},
  \href{https://doi.org/10.1088/1475-7516/2020/04/027}{\emph{Journal of
  Cosmology and Astroparticle Physics} {\bfseries 2020} (2020) 27}
  [\href{https://arxiv.org/abs/1912.06104}{{\ttfamily 1912.06104}}].

\bibitem{LindblomEtAl2008}
L.~Lindblom, B.J.~Owen and D.A.~Brown, \emph{Model waveform accuracy standards
  for gravitational wave data analysis},
  \href{https://doi.org/10.1103/physrevd.78.124020}{\emph{Physical Review D}
  {\bfseries 78} (2008) } [\href{https://arxiv.org/abs/0809.3844}{{\ttfamily
  0809.3844}}].

\bibitem{GWTC4}
A.G.~Abac and others {(LIGO Scientific, Virgo, and KAGRA Collaborations)},
  ``{{GWTC-4}}.0: {{Updating}} the gravitational-wave transient catalog with
  observations from the first part of the fourth {{LIGO-Virgo-KAGRA}} observing
  run.'' Aug., 2025.

\bibitem{GWToolbox}
S.-X.~Yi, G.~Nelemans, C.~Brinkerink, Z.~Kostrzewa-Rutkowska, S.T.~Timmer,
  F.~Stoppa et~al., \emph{{The Gravitational Wave Universe Toolbox - A software
  package to simulate observations of the gravitational wave universe with
  different detectors,}},
  \href{https://doi.org/10.1051/0004-6361/202141634}{\emph{Astron. Astrophys.}
  {\bfseries 663} (2022) A155}
  [\href{https://arxiv.org/abs/2106.13662}{{\ttfamily 2106.13662}}].

\bibitem{NarikawaEtAl2015}
T.~Narikawa, K.~Ueno, H.~Tagoshi, T.~Tanaka, N.~Kanda and T.~Nakamura,
  \emph{Detectability of bigravity with graviton oscillations using
  gravitational wave observations},
  \href{https://doi.org/10.1103/PhysRevD.91.062007}{\emph{Physical Review D}
  {\bfseries 91} (2015) 062007}.

\bibitem{EastherEtAl2003}
R.~Easther, D.~Langlois, R.~Maartens and D.~Wands, \emph{Evolution of
  gravitational waves in {{Randall-Sundrum}} cosmology},
  \href{https://doi.org/10.1088/1475-7516/2003/10/014}{\emph{Journal of
  Cosmology and Astroparticle Physics} {\bfseries 2003} (2003) 014}
  [\href{https://arxiv.org/abs/hep-th/0308078}{{\ttfamily hep-th/0308078}}].

\bibitem{VerasEtAl2015}
D.F.S.~Veras, J.E.G.~Silva, W.T.~Cruz and C.A.S.~Almeida, \emph{Gravitational
  {{Kaluza-Klein}} modes in the string-cigar braneworld},
  \href{https://doi.org/10.1103/PhysRevD.91.065031}{\emph{Physical Review D}
  {\bfseries 91} (2015) 065031}
  [\href{https://arxiv.org/abs/1409.3180}{{\ttfamily 1409.3180}}].

\bibitem{Planck2018}
{\scshape Planck Collaboration} collaboration, \emph{Planck 2018 results:
  {{VI}}. {{Cosmological}} parameters},
  \href{https://doi.org/10.1051/0004-6361/201833910}{\emph{Astronomy \&
  Astrophysics} {\bfseries 641} (2020) A6}
  [\href{https://arxiv.org/abs/1807.06209}{{\ttfamily 1807.06209}}].

\bibitem{CembranosEtAl2017}
J.A.R.~Cembranos, A.L.~Maroto and H.~{Villarrubia-Rojo}, \emph{Constraints on
  hidden gravitons from fifth-force experiments and stellar energy loss},
  \href{https://doi.org/10.1007/JHEP09(2017)104}{\emph{Journal of High Energy
  Physics} {\bfseries 2017} (2017) 104}
  [\href{https://arxiv.org/abs/1706.07818}{{\ttfamily 1706.07818}}].

\bibitem{CembranosEtAl2022}
J.A.R.~Cembranos, R.L.~Delgado and H.~{Villarrubia-Rojo}, \emph{{{LHC}}
  constraints on hidden gravitons},
  \href{https://doi.org/10.1007/JHEP01(2022)129}{\emph{Journal of High Energy
  Physics} {\bfseries 2022} (2022) 129}
  [\href{https://arxiv.org/abs/2108.00930}{{\ttfamily 2108.00930}}].

\end{thebibliography}\endgroup

\end{document}